\input ./style/arxiv-general.cfg
\documentclass[MSNbibl,nameyear,dvips]{arxstspdf}
\makeatletter
   \@ifpackageloaded{graphicx}{}{\usepackage{graphicx}}
\makeatother
\usepackage{flushend}
\usepackage{stfloats}
\volume{30}
\issue{2}
\pubyear{2015}
\firstpage{176}
\lastpage{180}
\doi{10.1214/15-STS518}
\referstodoi{10.1214/14-STS487}
\docsubty{FLA}

\makeatletter
\newcommand{\rrvert}{\vert}
\newcommand{\llvert}{\vert}
\newcommand{\s}{\mathbf{s}}
\def\ss{\mathbf{s}}
\def\hh{\mathbf{h}}
\def\YY{\mathbf{Y}}
\def\ss{\mathbf{s}}
\def\kk{\mathbf{k}}
\makeatother

\begin{document}
\begin{frontmatter}
\vspace*{12pt}
\title{When Doesn't Cokriging Outperform~Kriging?}
\runtitle{Comment}

\begin{aug}
\author[A]{\fnms{Hao}~\snm{Zhang}\corref{}\ead[label=e1]{zhanghao@purdue.edu}}
\and
\author[B]{\fnms{Wenxiang}~\snm{Cai}}
\runauthor{H. Zhang and W. Cai}

\affiliation{Purdue University and The University of International
Business and Economics}

\address[A]{Hao Zhang is Professor, Department of Statistics and Department of Forestry and
Natural Resources, Purdue University, West Lafayette, Indiana 47907, USA \printead{e1}.}
\address[B]{Wenxiang Cai is Graduate Student, School of International Trade and
Economics, The University of International Business and Economics,
Beijing, China.}
\end{aug}

%
\begin{abstract}
Although cokriging in theory should yield smaller or equal prediction
variance than kriging,
this outperformance sometimes is hard to see in practice. This should
motivate theoretical studies
on cokriging. In general, there is a lack of theoretical results for cokriging.
In this work, we provide some theoretical results to compare cokriging with
kriging by examining some explicit models and specific sampling schemes.
\end{abstract}

%
\end{frontmatter}

\citet{GentonKleibercross} provided an excellent review of recent
development in the mutivariate covariance functions. In many
situations, the ultimate objective of modeling the multivariate
covariance function is to obtain superior prediction through cokriging.
In theory, cokriging should have a prediction variance no larger than
that of the kriging prediction. However, as the authors point out in
the paper, sometimes the improvement of cokriging is very little or
none. In this note, we try to shed some light through some theoretical
investigations.

For univariate Gaussian stationary processes, we now have a good
understanding of the properties of kriging and statistical inferences.
For example, theoretical results have been established to justify (i)
that two different covariance functions may yield asymptotically
equally optimal prediction \citep{Stein}, and (ii) some parameters are
not consistently estimable if the spatial domain is bounded \citep
{ZhangJASA04}.
We know the conditions under which a misspecified covariance function
yields an asymptotically right prediction and can exploit this fact to
simplify computations (\citeauthor{ZhangJASA04}, \citeyear{ZhangJASA04}; \citeauthor{DuZhangMandrekarTapering}, \citeyear{DuZhangMandrekarTapering}).

We lack the analogous understanding for the multivariate spatial
models. There are no explicit theoretical results to answer the
following questions:
\begin{itemize}
\item How important is the cross-covariance function? Specifically,
could two different multivariate covariance functions yield an
asymptotically equally optimal prediction?
\item Which parameters are important to cokriging? We know which
parameters are important to kriging.
\item How much improvement does cokriging have over kriging?
\end{itemize}

One particular concept that has been shown useful in the study of
kriging is the equivalence of probability measures due to a theorem
established by \citet{BlackwellDubins65}. Let $\s_i$, $i=1, \ldots,
n$ be sampling sites on a fixed domain (area) where
the process $Y(\s)$ is observed, and $\{\s_i, i> n\}$ be a set of
sites on the same domain where $Y$ is to be predicted. If the two
Gaussian measures $P_1$ and $P_2$ are equivalent on the $\sigma
$-algebra generated by $Y(\s_i), i=1, 2, \ldots,$ then with
$P_1$-probability one,
\begin{eqnarray*}
&& \sup\bigl\llvert P_1 \bigl\{A|Y(\s_i), i=1, \ldots, n
\bigr\}
\\
&&\qquad{} - P_2 \bigl\{A|Y(\s_i), i=1, \ldots, n \bigr\}
\bigr\rrvert
\\
&&\quad \to0\qquad\mbox{as } n\to\infty,
\end{eqnarray*}
where the supremum is taken over $A\in
\sigma\{Y(\s_i), i>n\}$. The above result implies that the linear
predictions under the two measures are asymptotically equally optimal
\citep{Stein}.

This result can be readily extended to the multivariate spatial process
and therefore implies two cokriging predictors are asymptotically
equally optimal under the two probability measures if the two Gaussian
measures are equivalent. However, unlike in the univaritate case, there
are very limited results on equivalence of probability measures.
\citet
{PorcuEquiv} gave some general conditions for equivalent measures for
multivariate Gaussian processes though there is still a lack of
explicit examples where equivalent measures occur.

We now provide some sufficient conditions for the equivalent of
Gaussian measures for a particular bivariate model. Let $\YY(\s
)=(Y_1(\s), Y_2(\ss))'$ be a stationary bivariate
Gaussian process
with the following bivariate covariance function under the probability
measure $P_k$, $k=1, 2$, such that
\begin{eqnarray*}
C_{ij}(\hh) &=& \operatorname{Cov} \bigl(Y_i(\ss), Y_j(\ss+\hh)
\bigr)
\\
&=& M \bigl(\llvert\hh\rrvert, \sigma_{ij,k}, \alpha_{k},
\nu\bigr),\quad i, j=1, 2,
\end{eqnarray*}
where $M(\cdot,\sigma^2, \alpha, \nu)$ denotes the Mat\'{e}rn
covariance function with variance $\sigma^2$, scale parameter $\alpha
$ and the smoothness parameter $\nu$. The following are sufficient
conditions for the two measures $P_k$ to be equivalent on the $\sigma
$-algebra generated by $\{Y_i(\ss), \ss\in D, i=1, 2\}$ for some
bounded set $D\in R^d$, $d\le3$:
%
%
\begin{eqnarray}\label{eqnequiv}
\sigma_{ii,1}^2 \alpha_{1}^{2\nu}&=&
\sigma_{ii,2}^2 \alpha_{2}^{2\nu},
\nonumber\\[-8pt]\\[-8pt]\nonumber
\sigma_{12,1}/\sqrt{\sigma_{11,1}\sigma_{22,1}}&=&\sigma
_{12,2}/\sqrt{\sigma_{11,2}\sigma_{22,2}}.
\end{eqnarray}

To prove this claim, we employ the Karhunen--Lo\`eve expansion under
measure $P_1$. Since the two processes $\{Y_i(\ss)/\sqrt{\sigma
_{ii,1}}\}$, $i=1, 2$, have the same covariance function $M(|\hh|,a,
\alpha, \nu)$ and therefore possess the same Karhunen--Lo\`eve
expansion under measure $P_1$,
\[
\frac{Y_i(\ss)}{\sqrt{\sigma_{ii,1}}}=\sum_{l=1}^\infty\sqrt{
\lambda_l}f_l(\ss)Z_{il},
\]
where for $i=1, 2$, $\{Z_{il}, l=1,\ldots\}$ consists of i.i.d.
standard normal random variables under measures $P_1$. Clearly, the
eigenvalues $\lambda_l$ and eigenfunctions $f_l(\ss)$ only depend on
the correlation function and hence do not depend on $i$. In addition,
\[
Z_{il}=\frac{1}{\sqrt{\lambda_l \sigma_{ii,1}}}\int_D
Y_i(\ss)f_l(\ss)\,d\ss.
\]

Using the above expression, it is not hard to show that
%
%
\begin{eqnarray}
E_1(Z_{1l}Z_{2m})&=&r \delta_{l,m}
\nonumber\\[-8pt]\label{eqn2}\\ [-8pt]
\eqntext{\mbox{for } r=\sigma_{12,1}/\sqrt{\sigma_{11,1}
\sigma_{22,1}},}
\\
E_2(Z_{1l}Z_{2m})&=&r E_2(Z_{1l}Z_{1m}).\label{eqn3}
\end{eqnarray}

The Karhunen--Lo\`eve expansion implies that\break $\{Z_{il}, l=1, 2, \ldots,
\infty\}$ is a basis of the Hilbert space generated by $\{Y_i(\ss),
\ss\in D\}$ with respect to measure $P_1$. Hence, $\{Z_{1l}, Z_{2l},
l=1, 2,\ldots \}$ is a basis of the Hilbert space generated by the
two processes $\{Y_i(\ss), i=1, 2, \ss\in D\}$. The two measures are
equivalent on the Hilbert space if and only if they are so on $\{
Z_{1i}, Z_{2i}, i=1, 2,\ldots \}$ (\citeauthor{Ibragimov78}, \citeyear{Ibragimov78}, page~72). To
show the equivalence of the two measures, we only need to verify (\citeauthor{Stein}, \citeyear{Stein}, page~129)
%
%
\begin{eqnarray}\label{eqnproof}
&& \sum_{i=1}^2\sum
_{j=1}^2\sum_{l=1}^\infty
\sum_{m=1}^\infty\bigl(E_1(Z_{il}Z_{jm})-E_2(Z_{il}Z_{jm})
\bigr)^2
\nonumber\\[-8pt]\\[-8pt]\nonumber
&&\quad <\infty.
\end{eqnarray}
Because conditions (\ref{eqnequiv}) imply that the two measures are
equivalent on $\{Y_i(\ss), \ss\in D\}$ \citep{ZhangJASA04}, we must have
\[
\sum_{l=1}^\infty\sum
_{m=1}^\infty\bigl(E_1(Z_{il}Z_{im})-E_2(Z_{il}Z_{im})
\bigr)^2<\infty,\quad i=1, 2.
\]
For $i\ne j$, equations (\ref{eqn2}) and (\ref{eqn3}) imply
\begin{eqnarray*}
&& \sum_{l=1}^\infty\sum
_{m=1}^\infty\bigl(E_1(Z_{1l}Z_{2m})-E_2(Z_{1l}Z_{2m})
\bigr)^2
\\
&&\quad =r^2 \sum_{l=1}^\infty
\sum_{m=1}^\infty\bigl(E_1(Z_{1l}Z_{1m})-E_2(Z_{1l}Z_{1m})
\bigr)^2<\infty.
\end{eqnarray*}
Therefore, (\ref{eqnproof}) is proved and so is the sufficiency of
the conditions. We now have an explicit example where two different
bivariate covariance functions yield asymptotically equal cokriging results.

Next, we will try to explain why sometimes it is hard to see the
improvement of cokriging over the kriging prediction. Consider a
bivariate Gaussian process with mean 0 and exponential covariance
functions such that
%
%
\begin{eqnarray}\label{eqn4*}
C_{ij}(\hh)&=&\operatorname{Cov} \bigl(Y_i(\ss), Y_j(\ss+\hh)
\bigr)
\nonumber\\[-8pt]\\[-8pt]\nonumber
&=&\sigma_{ij}\exp\bigl(-\alpha\llvert\hh\rrvert\bigr),\quad i, j=1,2.
\end{eqnarray}
Assume the two processes are observed at $n$ points $\ss_i, i=1,
\ldots, n$, and predict $Y_1(0)$. Write $\YY_1=(Y_1(\ss_i), i=1,
\ldots, n)'$,
$\YY_2=(Y_2(\ss_i), i=1, \ldots, n)'$. It is known that in this case
the cokriging predictor is identical to the kriging predictor. To see
this, let $R$ denote the correlation matrix of $\YY_1$, which is also
the correlation matrix of $\YY_2$. Then
\[
\operatorname{Cov}(\YY_i, \YY_j)=\sigma_{ij} R.
\]
Let $V$ be the matrix with $(i, j)$th element $\sigma_{ij}$. Then the
covariance matrix of $(\YY_1, \YY_2)$ is $V\otimes R$.
Let $\kk$ denote the vector of correlation coefficients between
$Y_1(\ss)$, the variable to be predicted, and $\YY_1$. Then
%
%
\begin{eqnarray}
&& E \bigl(Y_1(\ss)|\YY_1, \YY_2 \bigr)
\nonumber\\[-8pt]\\[-8pt]\nonumber
&&\quad =
\bigl((\sigma_{11}, \sigma_{22})\otimes\kk'
\bigr) \bigl(V^{-1}\otimes R^{-1} \bigr)\YY
\\
&&\quad = \bigl( \bigl(\kk', 0 \bigr)\otimes R^{-1} \bigr) \YY=\kk' R^{-1}\YY_1
\nonumber\\[-8pt] \label{eqn6}\\[-8pt]\nonumber
&&\quad =E \bigl(Y_1(\ss)|
\YY_1 \bigr).
\end{eqnarray}
Therefore, cokriging is identical to kriging and we should not expect
any improvement of cokriging over kriging. We can also show that they
are identical if $Y_2(s)$ is observed at a subset of locations where
$Y_1$ is observed.

One scenario where cokriging might outperform kriging is when the
auxiliary variable is observed at more locations than the predicted
variable. In the next example, we will examine analytically what
variables affect the improvement of cokriging over kriging. We assume
the same bivariate model (\ref{eqn4*}) and $Y_2(s)$ are observed at
$s\in O=\{i/n, i=\pm1, \pm2, \ldots, \pm n\}$, but $Y_1(s)$ is
observed at half of the points $s\in O_1=\{2i/n, i=\pm1, \pm2, \ldots
, \pm n/2\}$ where $n$ is an even integer. Denote the kriging predictor
and cokriging predictor of $Y_1(0)$ by
%
%
\begin{eqnarray}
\quad \hat Y_1(0)&=&E \bigl(Y_1(0)|Y_1(s), s\in
O_1 \bigr),
\\
\tilde Y_1(0)&=&E \bigl(Y_1(0)|Y_1(s), s
\in O_1, Y_2(t), t\in O \bigr).
\end{eqnarray}
We will derive the following asymptotic relative efficiency of kriging
to cokriging:
%
%
\begin{equation}
\lim_{n\to\infty} \frac{E(Y_1(0)-\tilde Y_1(0))^2}{E(Y_1(0)-\hat
Y_1(0))^2}=1-r^2/2,\label{eqn9}
\end{equation}
where $r$ is the correlation coefficient of $Y_1(s)$ and $Y_2(s)$.

The asymptotic relative efficiency of kriging prediction does not
depend on the scale parameter $\alpha$. Intuitively this is
understandable. However, for a finite sample size $n$, $\alpha$ may
affect the efficiency. We now present a simulation study to see how
$\alpha$ and $r$ affect the relatively efficiency of kriging
prediction. We consider the exponential covariance model with $\sigma
_{11}=\sigma_{22}=1$ and $r=0.2$ and $0.5$, and $\alpha=2, 4$ and
$8$. The auxiliary variable $Y_2$ is observed at $\pm i/n$, $i=1,
\ldots, n$, but the primary variable $Y_1$ is observed at $\pm i/n$
for even integers $0<i\le n$. We calculate the prediction variance for
predicting $Y_1(0)$ using both kriging and cokriging and obtain the
relative efficiency of kriging for different $n$, $\alpha$ and $r$.

%
\begin{figure*}

\includegraphics{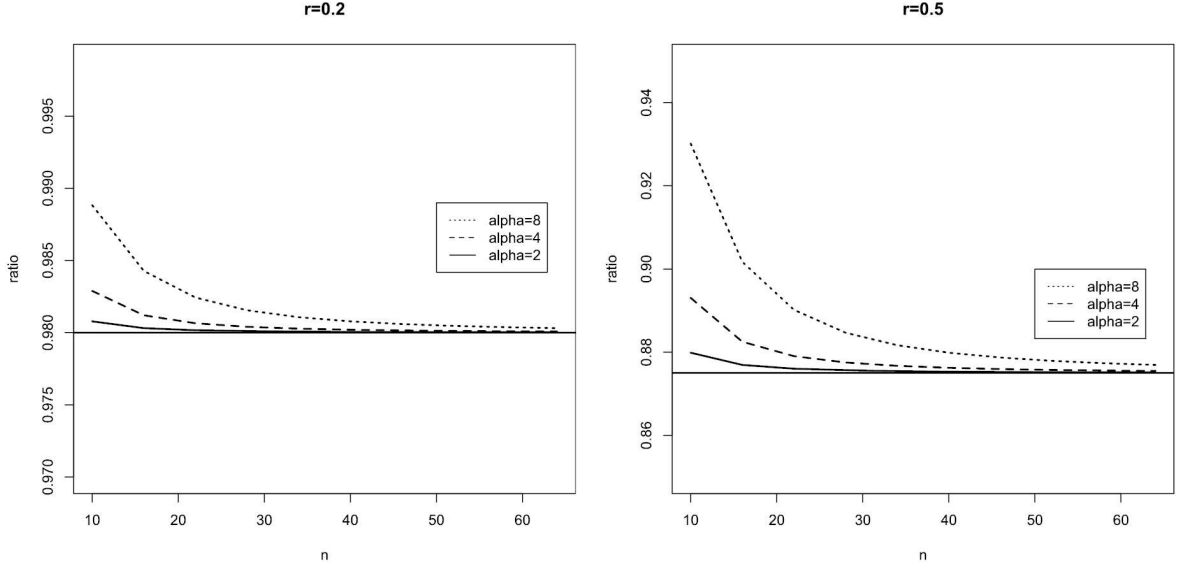}

\caption{Relative efficiency of kriging to cokriging for different
$r$, $\alpha$ and $n$. The solid horizonal line is the asymptotic
relative efficiency $1-r^2/2$.}\label{Figure1}
\end{figure*}

Figure~\ref{Figure1} plots the relative efficiency for different $r$,
$\alpha$ and $n$. We see that the relative efficiency of kriging
decreases as $n$ increases, which means that it is more likely to see
the outperformance of cokriging over kriging when $n$ is larger. When
the spatial autocorrelation is strong (i.e., $\alpha$ smaller), the
asymptotic efficiency is achieved relatively faster (i.e., with $n$ not
too larger). This agrees with many other infill asymptotic results.

We now prove (\ref{eqn9}). We first note a Markovian property of the
exponential model established by Du, Zhang and Mandrekar (\citeyear{DuZhangMandrekarTapering}),
which says $E(Y_1(s)|Y_1(s), s\in B) $ only depends on the two nearest
neighbors of $s$ in a finite set $B$ such that $s$ is between the
minimum and the maximum elements of $B$ (\citeauthor{DuZhangMandrekarTapering}, \citeyear{DuZhangMandrekarTapering}, Lemma~1). Also from the lemma, we obtain
\[
E \bigl(Y_1(0)-\hat Y_1(0) \bigr)^2=2
\sigma_{11}^2\alpha/n+o \bigl(n^{-2} \bigr).
\]
In the extreme case when $r=1$, we can view the process $Y_1(s)$ being
observed at $O$. Then in this extreme case, the above equation implies
\[
E \bigl(Y_1(0)-\tilde Y_1(0) \bigr)^2=
\sigma_{11}^2\alpha/n+o \bigl(n^{-2} \bigr).
\]
The ratio in (\ref{eqn9}) is clearly $1/2$. Hence, we have verified~(\ref{eqn9}) for this extreme case. On the other hand, when $r=0$,
the two predictors $\hat Y_1(0)$ and $\tilde Y_1(0)$ are identical and
(\ref{eqn9}) is obviously true.

We are going to show that
%
%
\begin{eqnarray}\label{eqn10}
\tilde Y_1(0)&=&b_1Y_1(-2/n)+b_2
Y_1(2/n)
\nonumber
\\
&&{}+ b_3 Y_2(-2/n)+b_4 Y_2(-1/n)
\\
&&{}+b_5 Y_2(1/n)+b_6Y_2(2/n),\nonumber
\end{eqnarray}
where
%
%
\begin{eqnarray}
b_1&=&b_2=\frac{{\mathrm e}^{-2\alpha/n}}{{\mathrm e}^{-4\alpha/n}+1},
\nonumber\\[-8pt] \label{eqnb12} \\[-8pt]\nonumber
b_3&=& b_6=-\frac{r{{\mathrm e}^{-2\alpha/n}}}{{\mathrm e}^{-4\alpha/n}+1},
\\
b_4&=&b_5=\frac{r{\mathrm e}^{-\alpha/n}}{{\mathrm e}^{-2\alpha/n}+1}. \label{eqnb45}
\end{eqnarray}
Some straightforward calculation yields
\begin{eqnarray*}
&& E \bigl(Y_1(0)-\tilde Y_1(0) \bigr)^2
\\
&&\quad =-
\sigma_{11}^2
\bigl(-2 {{\mathrm
e}^{-4\alpha/n}}{r}^{2}+{{\mathrm e}^{-6\alpha/n}}+2 {{\mathrm e}^{-
2\alpha/n}}{r}^{2}
\\
&&\hspace*{101pt}{} +{{\mathrm e}^{-4\alpha/n}}-{{\mathrm e}^{-2\alpha/n}}-1\bigr)
\\
&&\hspace*{8pt}\qquad{} / \bigl(\bigl( {{\mathrm e}^
{-4\alpha/n}}+1 \bigr) \bigl( {{\mathrm e}^{-2\alpha/n}}+1 \bigr) \bigr)
\\
&&\quad =\sigma_{11}^2 \bigl(2-r^2 \bigr)\alpha/n+o
\bigl(n^{-2} \bigr).
\end{eqnarray*}

Then (\ref{eqn9}) immediately follows. Hence, it is sufficient to
show (\ref{eqn10}). It is possible to show that $Y_1(0)-\tilde
{Y}_1(0)$ is uncorrelated with any $Y_1(s), s\in O_1$ and with any
$Y_2(t), t\in O$. Hence, $\tilde{Y}_1(0)$ must be the best linear
prediction. Here we take an alternative but more intuitive approach. We
will apply the Markovian property of the Gaussian exponential model to
show that $\tilde{Y}_1(0)$ only depends on $Y_1(-2/n)$, $Y_1(2/n)$,
$Y_2(-2/n)$, $Y_2(-1/n)$, $Y_2(1/n)$ and $Y_2(2/n)$. Consequently, the
coefficients $b_i$'s in (\ref{eqnb12}) and (\ref{eqnb45}) can be
found by solving linear equations.

For any odd integer $i$ between $-n$ and $n$,
%
%
\begin{eqnarray}
\quad&& E \bigl(Y_2(i/n)|Y_1(s), s\in O_1,
Y_2(t), t\in O, t\ne i/n \bigr)
\nonumber
\\
&&\quad=E \bigl\{E \bigl(Y_2(i/n)|Y_1(t),
Y_2(t), t\in O, t\ne i/n \bigr) |\nonumber
\\
&&\hspace*{84pt} Y_1(s), s\in
O_1, t\in O, t\ne i/n \bigr\}
\nonumber\\[-8pt]\\[-8pt]\nonumber
&&\quad=E \bigl\{E \bigl(Y_2(i/n)|Y_2(t), t\in O, t\ne i/n
\bigr) |
\\
&&\hspace*{55pt} Y_1(s), s\in O_1, t\in O, t\ne i/n \bigr\}
\nonumber
\\
&&\quad=E \bigl\{ Y_2(i/n)|Y_2(t_{i-}),
Y_2(t_{i+}) \bigr\},\nonumber
\end{eqnarray}
where $t_{i-}$ and $t_{i+}$ are the two nearest neighbors of $i/n$ in
$O$. For example, for $i=-1$, $t_{i-}=-2/n$ and $t_{i+}=1/n$.

Define $e_i=Y_2(i/n)-E\{ Y_2(i/n)|Y_2(t_{i-}), Y_2(t_{i+})\}$ for an
odd $i$. Then $e_i$ is independent of $Y_1(s), s\in O_1$ and $Y_2(t),
t\in O$ and $t\ne i/n$. Consequently,
%
%
\begin{eqnarray}\label{eqn14}
\quad &&E \bigl(Y_1(0)|Y_1(s), s\in O_1,
Y_2(t), t\in O \bigr)
\nonumber
\\
&&\quad =E \bigl(Y_1(0)|Y_1(s), Y_2(s), s\in
O_1, e_i, i \mbox{ odd} \bigr)
\nonumber\\[-8pt]\\[-8pt]\nonumber
&&\quad =E \bigl(Y_1(0)|Y_1(s), Y_2(s), s\in
O_1 \bigr)
\\
&&\qquad{} +E \bigl(Y_1(0)| e_i, i\mbox{ odd} \bigr). \nonumber
\end{eqnarray}
The first term in the above equation depends only on $Y_1(-2/n)$ and
$Y_1(2/n)$ due to the Markovian property. For the second term, because
the cross-covariance function is proportional to the covariance
function of $Y_2(t)$, we have
\[
E \bigl(Y_1(0)| e_i, i \mbox{ odd} \bigr)=r E
\bigl(Y_2(0)| e_i, i \mbox{ odd} \bigr).
\]
Applying again the property of conditional expectation and the
Markovian property, we get
\begin{eqnarray}
&&E \bigl(Y_2(0)| e_i, i \mbox{ odd} \bigr)
\nonumber
\\
&&\quad=E \bigl(E \bigl\{Y_2(0)|Y_2(t), t\in O \bigr\}|
e_i, i \mbox{ odd} \bigr)
\nonumber
\\
&&\quad=\beta E \bigl\{Y_2(-1/n)+Y_2(1/n)|e_i, i
\mbox{ odd} \bigr\}
\nonumber
\\
&&\quad=\beta E \bigl\{Y_2(-1/n)+Y_2(1/n)|e_{-1},
e_1 \bigr\},
\nonumber
\end{eqnarray}
where $\beta$ is the constant in $E(Y_2(0)|Y_2(-1/n),\break  Y_2(1/n))=\beta
(Y_2(-1/n)+Y_2(1/n) ) $, and the last equation follows the fact that
$e_i$ is independent to $Y(1/n)$ and $Y_2(-1/n)$ if $i\ne1$ or $-$1.
Therefore, the second term of (\ref{eqn14}) is a linear function of
$e_{-1}$ and $e_1$ and hence a linear function of $Y_2(i/n)$, $i=-2, -1,
1$ and 2.

%


\section*{Acknowledgment}
Hao Zhang is supported by a Grant from the China
Social Science Foundation (11\&ZD167) and NSF Grant (IIS-1028291). Wenxiang Cai
is supported by the Excellent Dissertation Fund of the University of
International Business and Economics.


%

%
\end{document}